\documentclass[smus]{snow2e}
\input epsf
\begin{document}

\def\Emiss{\rlap{E}{/}}
\def\Dzero{D$\emptyset$}
\def\GeV{{\rm GeV}}
\def\lsim{\, \lower0.5ex\hbox{$\stackrel{<}{\sim}$}\, }
\def\gsim{\, \lower0.5ex\hbox{$\stackrel{>}{\sim}$}\, }

\title{Precision Measurements Of Heavy Objects Working Group Summary}

\setlength{\titleblockheight}{5.5cm}
\author{
Marcel Demarteau\rlap,$^a$
Vassilis Koulovassilopoulos\rlap,$^b$
Joseph Lykken\rlap,$^a$
Fredrick I. Olness\rlap,$^{c\dagger}$  \\
Stephen Parke\rlap,$^{a*}$
Randall J. Scalise\rlap,$^{c\dagger}$
Erich Varnes\rlap,$^d$
and G. P. Yeh\rlap,$^{a}$\thanks{%
Convenor, \ 
${}^\dagger$ sub-group report editor.
Work supported in part by NSF and DOE. 
}
\\~\\
${}^a$ \em Fermi National Accelerator Laboratory, Batavia, IL 60510 USA\\
${}^b$ \em Universitat de Barcelona, Dept. E.C.M., Facultat de Fisica,
        Diagonal 647, 08028 Barcelona, SPAIN\\
${}^c$ \em Southern Methodist University, Department of Physics,
      Dallas, TX 75275-0175 USA\\
${}^d$ \em Lawrence Berkeley National Laboratory, Berkeley, CA 94720 USA\\
}

\maketitle

\thispagestyle{empty}\pagestyle{empty}

\begin{abstract} 
We report on the activities of the Precision Measurements Of Heavy Objects 
working group of the Very Large Hadron Collider Physics and Detector Workshop.
\end{abstract}

\section{Introduction}

The topics discussed by the  Precision Measurements Of Heavy Objects
working group spanned a very wide range; consequently, it is impossible
to cover each topic in depth. 
 Therefore,  in this report we will primarily focus
on the issues most relevant to a VLHC machine. 
 In the following, we mention only the highlights, and
refer the reader to the literature for more specific questions.

\section{Parton Distributions for VLHC\protect\footnote{%
\lowercase{\uppercase{B}ased on the presentation by 
\uppercase{F}redrick \uppercase{O}lness.}}
}


\def\figFredi{
\begin{figure}[htbp]
\begin{center}
\leavevmode
 \epsfxsize=3in  \epsfbox{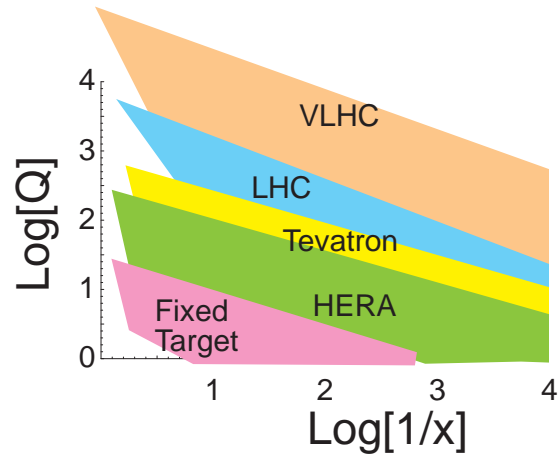}
\end{center}
\caption{Kinematic range of various machines.
Note the small $x$ range is clipped in this plot.  The $Q$ scale is in
GeV and the logs are base 10.
}
\label{fig:Fredi}
\end{figure}
}


\def\figFredii{
\begin{figure}[htbp]
\begin{center}
\leavevmode
 \epsfxsize=1.7in  \epsfbox{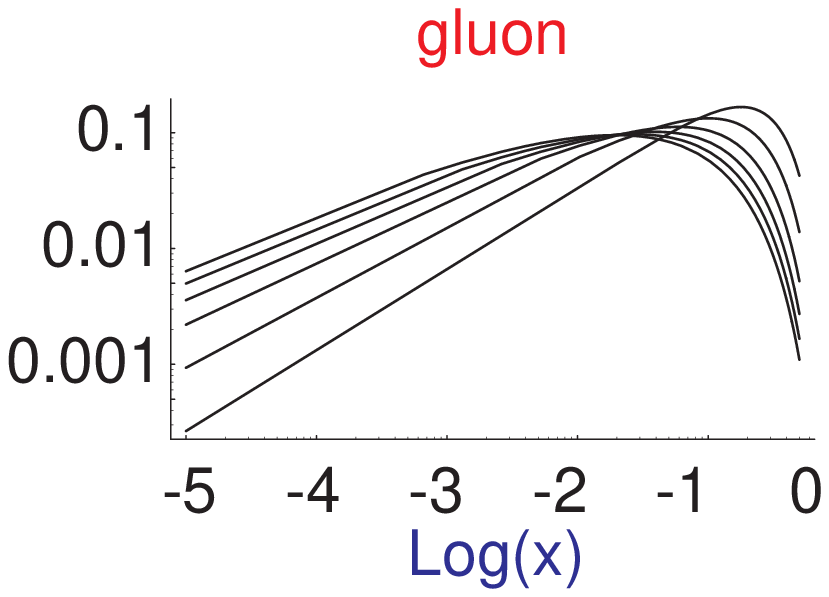}
 \epsfxsize=1.7in  \epsfbox{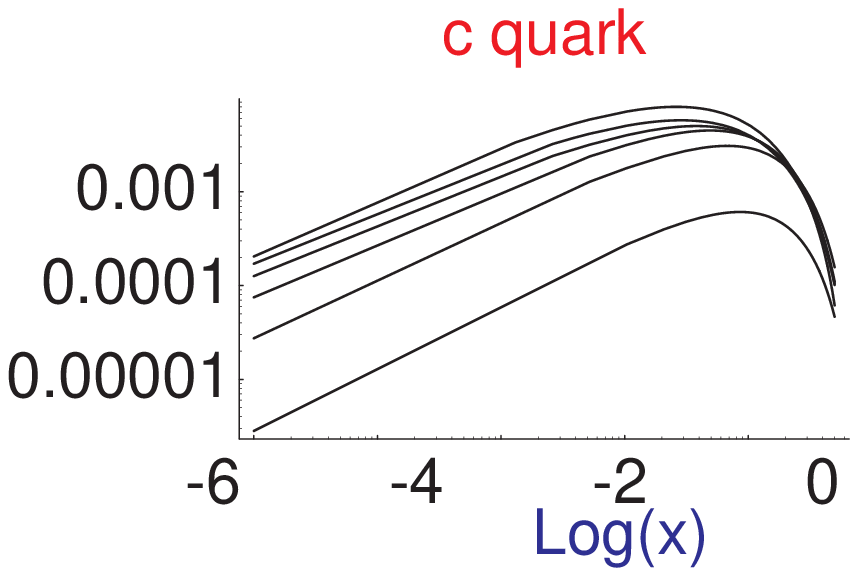}
\end{center}
\caption{Evolution of the a) gluon and b) charm PDF's
in $Q$ {\it vs.} $x$.
We display $x^2 f_{i/P}(x,Q)$ for 
$Q=\{2, 10^1,10^2,10^3,10^4,10^5 \}$ GeV.
}
\label{fig:Fredii}
\end{figure}
}


\def\figFrediii{
\begin{figure}[htbp]
\begin{center}
\leavevmode
 \epsfxsize=1.7in  \epsfbox{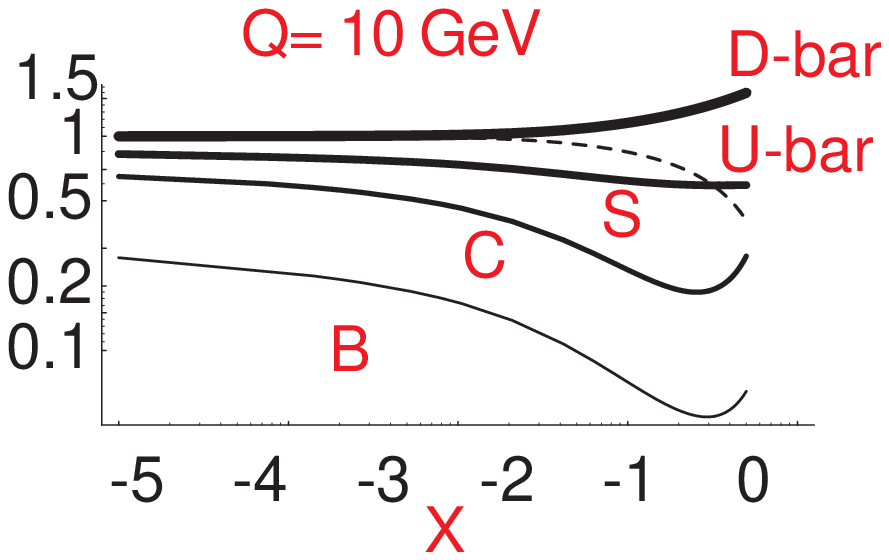}
 \epsfxsize=1.7in  \epsfbox{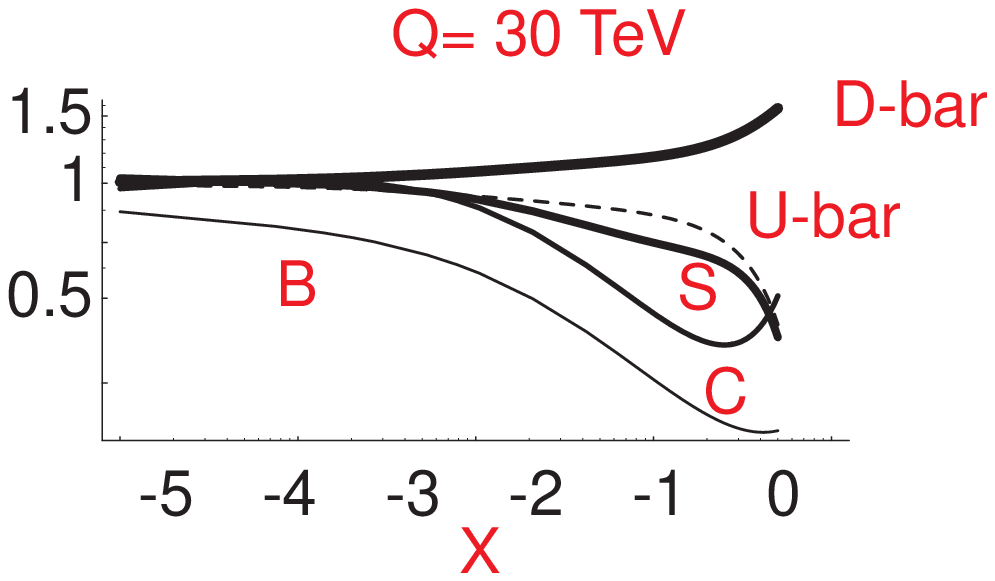}
\end{center}
\caption{Flavor democracy at a) 10 GeV and b) 30 TeV.
We compare the individual parton distributions $f_{i/P}(x,Q)$
to that of the average sea, $(\bar{u} + \bar{d})/2$.
}
\label{fig:Frediii}
\end{figure}
}

Global QCD analysis of lepton-hadron and hadron-hadron  processes has
made steady progress in testing the consistency of perturbative QCD
(pQCD) within many different sets of data, and in yielding
increasingly detailed information on the universal parton
 distributions.\footnote{PDF sets are available via WWW on the CTEQ page at
http://www.phys.psu.edu/$\sim$cteq/ and  on the The Durham/RAL HEP
Database at  http://durpdg.dur.ac.uk/HEPDATA/HEPDATA.html.}

 We present  the kinematic ranges covered by selected facilities
relevant for the determination of the universal parton distributions.
 While we would of course like to probe the full $\{x,Q\}$ space,
the small $x$ region is of
special interest.  For example, the rapid rise of the $F_2$ structure
function observed at HERA  suggests that we may reach the parton
density saturation region more quickly than anticipated.
Additionally, the small $x$ region can serve as a useful testing
ground for BFKL, diffractive phenomena, and similar processes.
Conversely, the production of new and exotic phenomena generally
happens in the region of relatively high $x$ and $Q$.

 This  compilation provides a useful guide to the planning of future
experiments and to the design of strategies for global analyses.
Another presentation regarding future and near-future machines
is given in the 1996 Snowmass Structure Functions Working Group 
report.\cite{sfun}

Here we will simply mention a few features which are particularly relevant
for such a very high energy facility as a VLHC.

\figFredi

As we see in Fig.~\ref{fig:Fredi}, the VLHC will probe an $\{ x,Q\}$ region 
far beyond the range of present data.  To accurately calculate processes
at a VLHC, we must have precise PDF's in this complete kinematic
range.  Determining the PDF's in the small $x$ regime is a serious
problem since there will be no other measurement in the extreme
kinematic domain required by VLHC.  For the large $x$ and $Q$ region, the
PDF's at large $Q$ can, in principle, be determined via the standard
QCD DGLAP evolution, but in practice uncertainties from the small $x$
region can contaminate this region.

\figFredii

\figFrediii

In Fig.~\ref{fig:Fredii}, we display the evolution of the PDF's for a 
selection of 
partons.  For the gluon and the valence quarks, we see a decrease at high $x$
and an increase at low $x$ with $x \sim 0.1$ as the crossing point.
In contrast, for the heavy quark PDF's, we see generally an increase 
with increasing $Q$.  The momentum
fraction of the partons {\it vs.} energy scale is shown in 
Table~\ref{tab:momfrac}.
An interesting feature to note here is the approximate ``flavor
democracy" at large energy scales; that is, as we probe the proton at
very high energies, the influence of the quark masses becomes
smaller, and all the partonic degrees of freedom carry comparable
momentum fractions.  To be more precise, we see that at the very
highest energy scales relevant for the VLHC, the strange and charm
quark are on par with the up and down sea, (while the bottom quark
lags behind a bit).  This feature is also displayed in Fig.~\ref{fig:Frediii}
where we show these contributions for two separate scales.  In light of this
observation, we must dispense with preconceived notions of what are
``traditionally" heavy and light quarks, and be prepared to deal with
all quark on an equal footing at a VLHC facility. This approach is
discussed in the following section.

\def\bigstrut{\vrule height16pt depth6pt width0pt}%
\begin{table}[htbp]
\begin{center}
\caption{Momentum fraction (in percent) carried by separate partons as a 
function of the energy scale $Q$.
}
\label{tab:momfrac}
\begin{tabular}{||r|c|c|c|c|c|c||} \hline \hline 
\bigstrut
Q \qquad & 	$g$	 & $\bar{u}$	 & $\bar{d}$	 & $s$	 & $c$	 
& $b$  \\ \hline 
  3 GeV & 46 & 5 & 7 & 3 & 1 & 0  \\ \hline 
 10 GeV & 48 & 6 & 8 & 4 & 2 & 0  \\ \hline 
 30 GeV & 48 & 6 & 8 & 5 & 3 & 1  \\ \hline 
100 GeV & 48 & 7 & 8 & 5 & 3 & 2  \\ \hline 
300 GeV & 49 & 7 & 8 & 6 & 4 & 2  \\ \hline 
  1 TeV & 49 & 7 & 8 & 6 & 4 & 3  \\ \hline 
  3 TeV & 49 & 7 & 8 & 6 & 4 & 3  \\ \hline 
 10 TeV & 50 & 7 & 9 & 6 & 5 & 4  \\ \hline 
 30 TeV & 50 & 7 & 9 & 7 & 6 & 4  \\ \hline 
100 TeV & 51 & 7 &10 & 7 & 7 & 4  \\ \hline  
\hline 
\end{tabular}
\end{center}
\end{table}

\section{Heavy Quark Hadroproduction\protect\footnote{%
\lowercase{\uppercase{B}ased on the presentation by 
\uppercase{R}andall \uppercase{J}. \uppercase{S}calise.}}
}


\def\figHQdata{
\begin{figure}[htbp]
\begin{center}
\leavevmode
 \epsfxsize=3in  \epsfbox{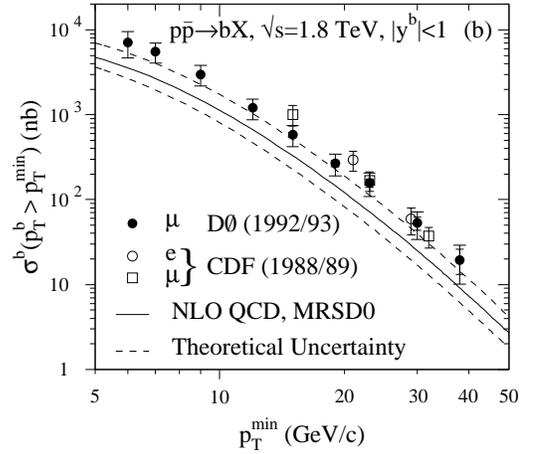}
\end{center}
      \caption{
Heavy quark hadroproduction data. 
{\it Cf.}, Ref.~\protect\cite{hqdata}.
}
   \label{fig:figHQdata}
\end{figure}
}

\def\figcacciari{
\begin{figure}[htbp]
\begin{center}
\leavevmode
 \epsfxsize=3in  \epsfbox{cacciari.eps}
\end{center}
      \caption{
Scale dependence of the heavy quark hadroproduction cross section
as a function of  $\mu = \xi \mu_{ref}$ at $y=0$ and $p_t= 80\, \GeV$. 
The NDE curve is the calculation of 
Ref.~\protect\cite{nde}.
The {\it fragm. funct.} and {\it born} curves are the calculation of
Ref.~\protect\cite{Greco}. }
   \label{fig:figcacciari}
\end{figure}
}

\def\figProd{
\begin{figure}[htbp]
\begin{center}
\leavevmode
\hbox{
 \epsfxsize=0.40\textwidth  \epsfbox{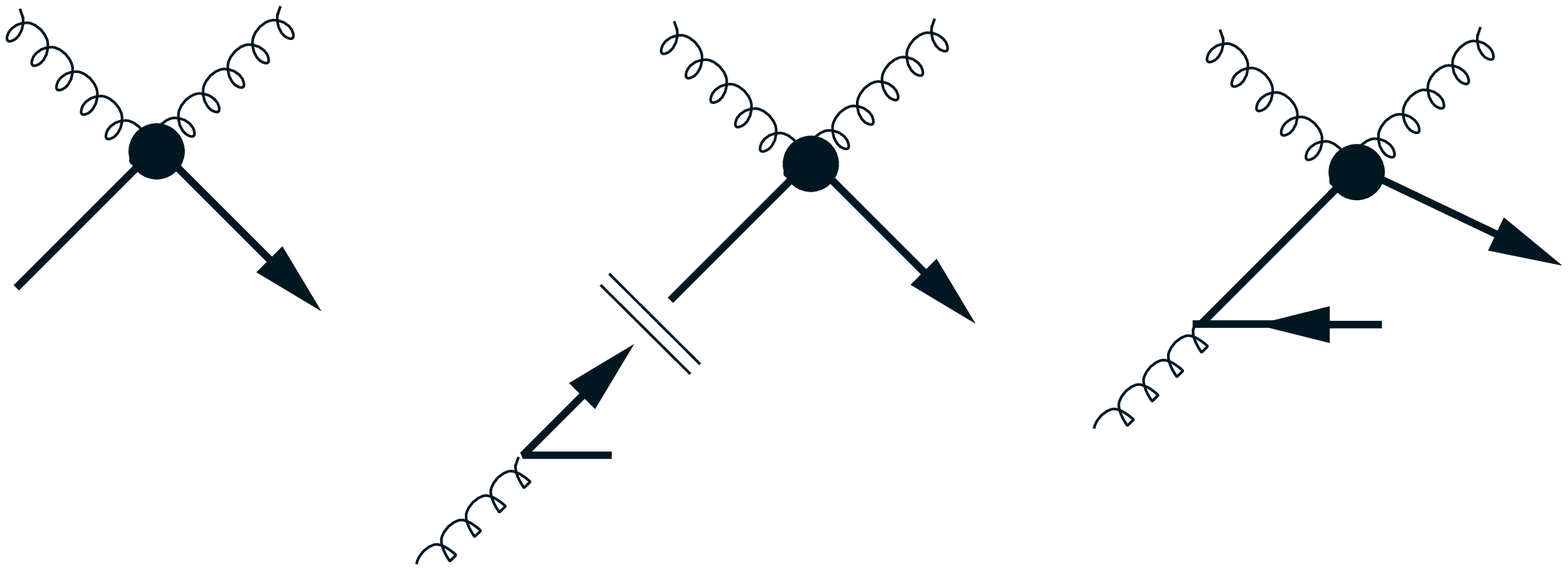}
}
\end{center}
      \caption{
a)~Generic leading-order diagram for heavy-flavor excitation (LO-HE), $gQ\to gQ$.
b)~Subtraction diagram for heavy-flavor excitation (SUB-HE),
   ${}^1f_{g\to Q} \otimes \sigma(gQ\to gQ)$.
c)~Next-to-leading-order diagram for heavy-flavor creation (NLO-FC).
\null\hfill\null}
   \label{fig:figProd}
\end{figure}
}

\def\figDecay{
\begin{figure}[htbp]
\begin{center}
\leavevmode
\hbox{
 \epsfxsize=0.40\textwidth  \epsfbox{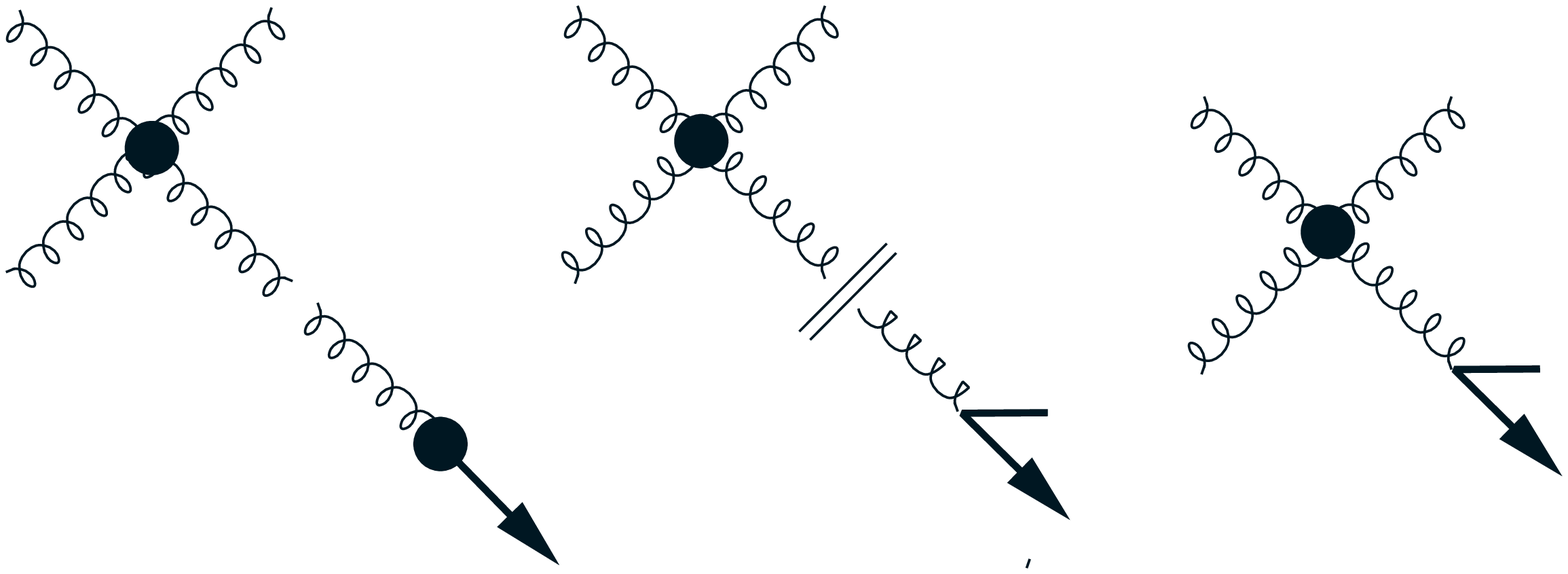}
}
\end{center}
      \caption{
a)~Generic leading-order diagram for heavy-flavor fragmentation (LO-HF),
   $\sigma(gg\to gg) \otimes D_{g\to Q}$.
b)~Subtraction diagram for heavy-flavor fragmentation (SUB-HF),
   $\sigma(gg\to gg) \otimes {}^1d_{g\to Q}$.
c)~Next-to-leading-order diagram for heavy-flavor creation (NLO-FC).
\null\hfill\null}
   \label{fig:figDecay}
\end{figure}
}
\def\figFeSub{
\begin{figure}[htbp]
\begin{center}
\leavevmode
 \hbox{
 \epsfxsize=0.22\textwidth  \epsfbox{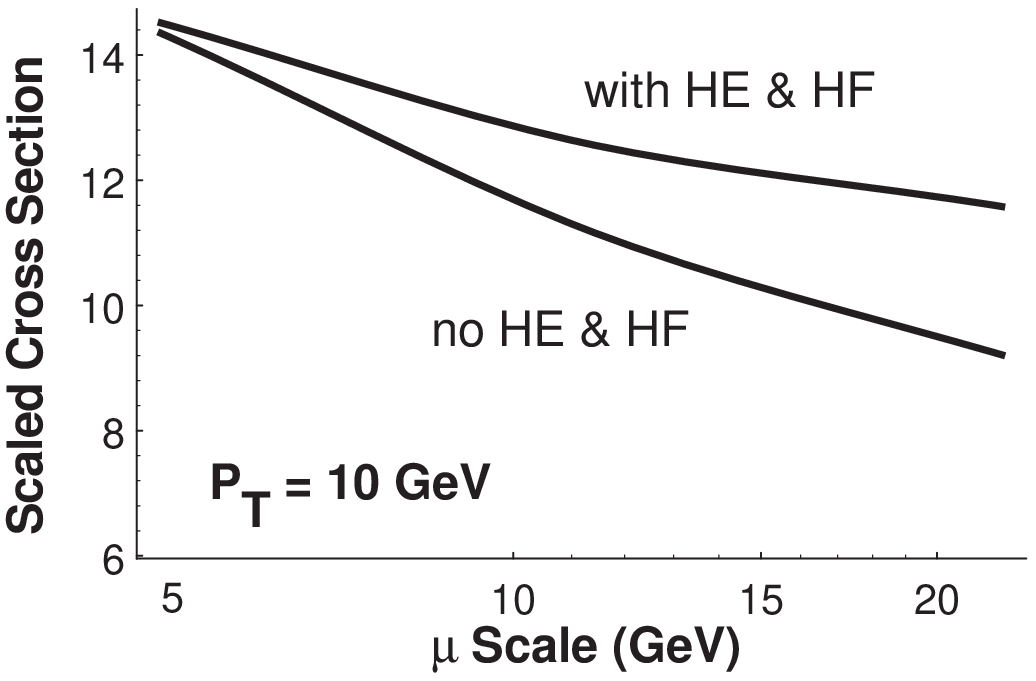}
 \hfill 
 \epsfxsize=0.22\textwidth  \epsfbox{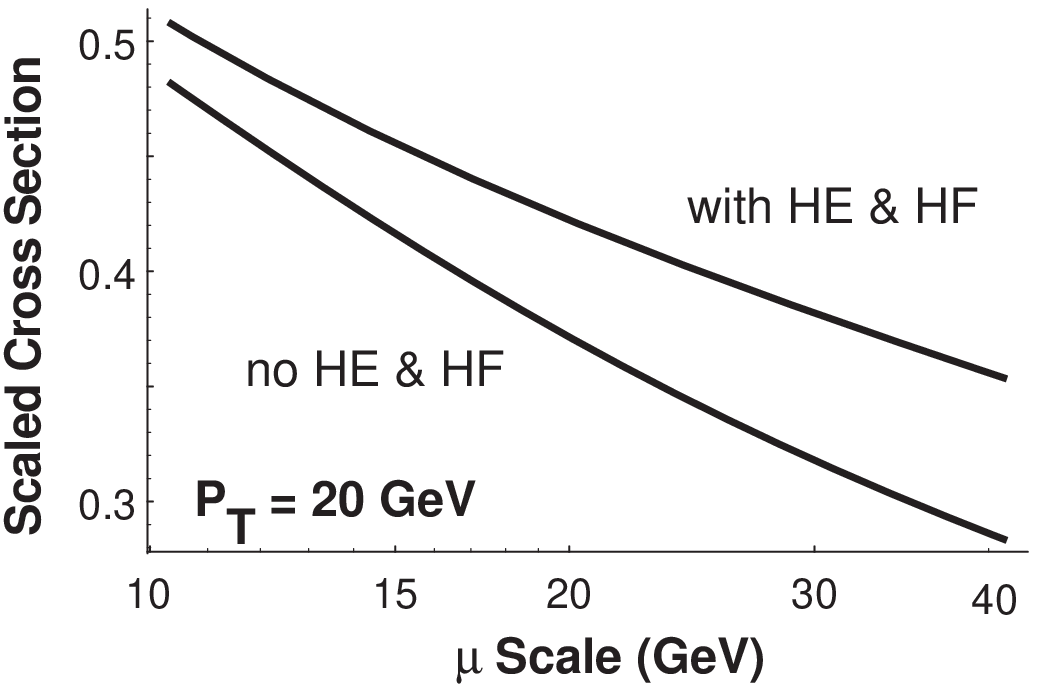}
}
\end{center}
      \caption{
The scaled differential cross section $p_T^5 \, d^2 \sigma/dp_T^2/dy$ at
$p_T=10, \, 20 \, \GeV$ and $y=0$ in  $(pb-\GeV^3)$ {\it vs.} $\mu$.
 The lower curves (thin line) are the heavy quark
production cross sections {\it ignoring}
heavy-flavor excitation (HE) and heavy-flavor fragmentation (HF).
 The upper curves (thick line) are the heavy quark
production cross sections {\it including}
HE  and HF. {\it Cf.}, Ref.~\protect\cite{cost}.
\null\hfill\null}
   \label{fig:figfesub}
\end{figure}
}


Improved experimental measurements of heavy quark hadroproduction has
increased the demand on the theoretical community for more precise
predictions.\cite{hqdata,nde,acot,Greco,cost} 
The first Next-to-Leading-Order (NLO)
calculations of charm and bottom hadroproduction cross sections were
performed some years ago.\cite{nde} As the accuracy of the data
increased, the theoretical predictions displayed some shortcomings:
 1) the theoretical cross-sections fell well short of the measured values,  
 and 
 2) they displayed a strong dependence on the unphysical renormalization
scale  $\mu$. 
Both these difficulties indicated that these predictions were missing
important physics. 

 \figHQdata
 \figProd 
 \figDecay 
 \figFeSub

These deficiencies can, in part, be traced to 
large contributions generated by logarithms associated with the heavy quark 
mass scale, such as\footnote{Here, $m_Q$ is the heavy quark mass, $s$ is
the energy squared, and $p_T$ is the transverse momentum.}
$\ln(s/m_Q^2)$ and $\ln(p_T^2/m_Q^2)$.  Pushing the calculation to one
more order, formidable as it is, would not necessarily improve the situation 
since these large logarithms persist to every order of perturbation theory.
Therefore, a new approach was required to include these logs.
 
In 1994, Cacciari and Greco\cite{Greco} observed that since the heavy
quark mass played a limited dynamical role in the high $p_t$ region,
one could instead use the massless NLO jet calculation convoluted with
a fragmentation into a massive heavy quark pair to compute 
more accurately the production cross section in the region $p_t \gg m_Q$. 
In particular, they find that the dependence on the renormalization scale
is significantly reduced.

A recent study\cite{cost} investigated using initial-state heavy quark
PDF's and final-state fragmentation functions to resum the large
logarithms of the quark mass.  The principle ingredient was to include
the leading-order heavy-flavor excitation (LO-HE) graph 
(Fig.~\ref{fig:figProd})
and the leading-order heavy-flavor fragmentation (LO-HF) graph
(Fig.~\ref{fig:figDecay}) in the traditional NLO heavy quark
calculation.\cite{nde} These contributions can not be added naively to
the ${\cal O}(\alpha_s^3)$ calculation as they would double-count
contributions already included in the NLO terms; therefore, a
subtraction term must be included to eliminate the region of phase
space where these two contributions overlap.  This subtraction term
plays the dual role of eliminating the large unphysical collinear logs
in the high energy region, and minimizing the renormalization scale
dependence in the threshold region.  The complete calculation
including the contribution of the heavy quark PDF's and fragmentation
functions 1) increases the theoretical prediction, thus moving it
closer to the experimental data, and 2) reduces the $\mu$-dependence
of the full calculation, thus improving the predictive power of the
theory. (Cf., Fig~\ref{fig:figfesub}.)

In summary, the wealth of data on heavy quark hadroproduction will
allow for precise tests of many different aspects of the theory,
namely radiative corrections, resummation of logs, and multi-scale 
problems.  Resummation of the large logs associated with the
mass is an essential step necessary to bring theory in agreement with
current experiments and to make predictions for the VLHC.

\section{W Mass Studies\protect\footnote{%
\lowercase{\uppercase{B}ased on the presentation by 
\uppercase{M}arcel \uppercase{D}emarteau.}}
}

\def\figMWMT
{
\begin{figure}[bht]
 \epsfxsize=\hsize
 \centerline{\epsfbox{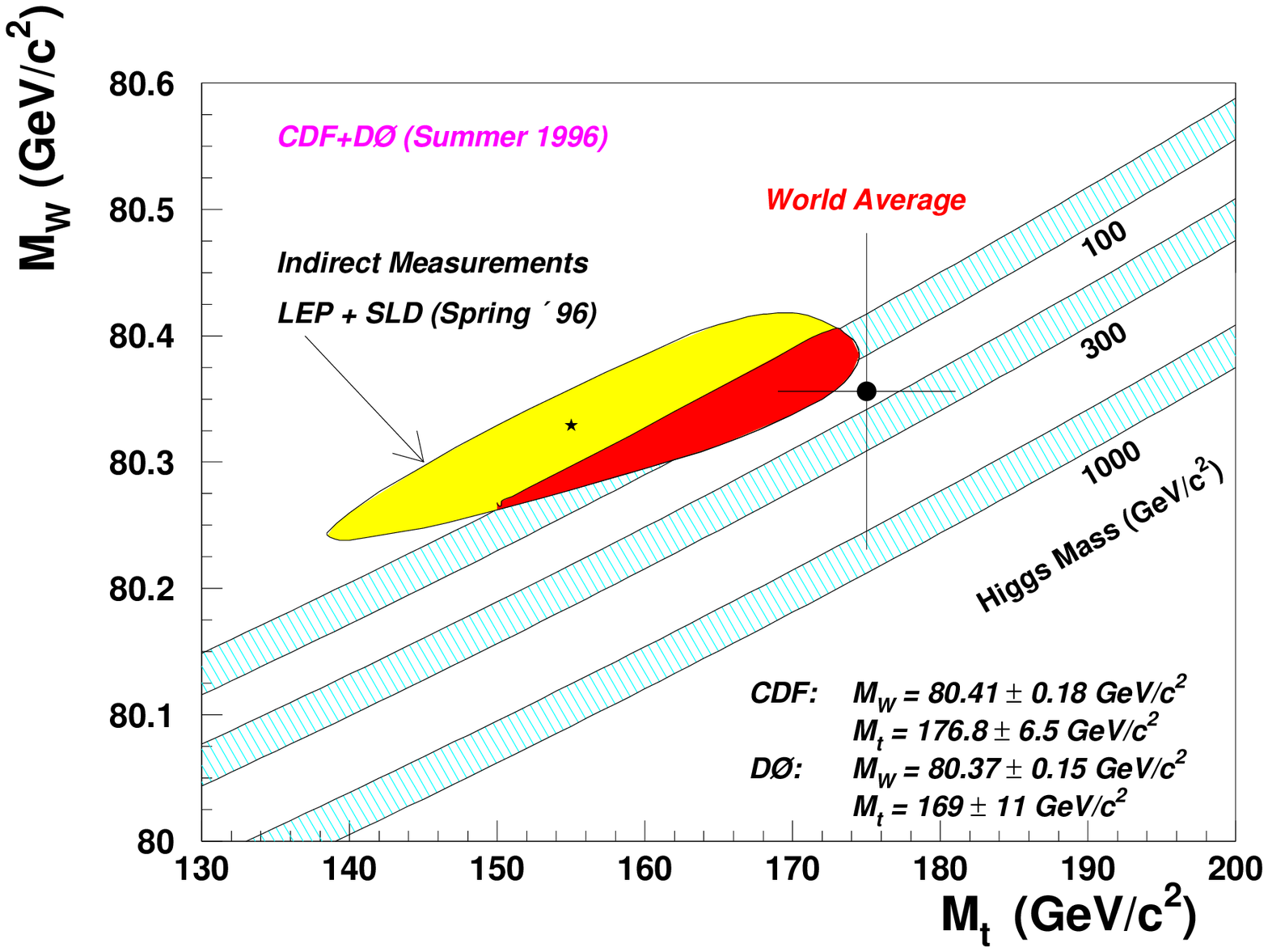}}
 \caption{
Plot of $M_W$ {\it vs.} $M_t$ with \Dzero\ and CDF preliminary measurements of
the W boson and top quark masses. Bands indicate the Standard Model
constraints for different Higgs mass values. Indirect measurements
from LEP I are also shown.
(June, 1997) 
Taken from Ref.~\protect\cite{baur_demarteau}.
}
 \label{fig:figMWMT}
\end{figure}
}
\figMWMT

The W boson mass is one of the fundamental parameters of the standard model;
its precision measurement can be used in conjunction with the top mass 
to extract information on the Higgs boson mass. 
 The W boson mass has already been measured precisely, and 
the current world average is:
$M_W = 80.356 \pm 0.125 \, \GeV/c^2$.

Here, we focus on issues which are unique to a VLHC facility, and
refer the reader to the literature for details regarding other 
topics.\cite{demarteau_354,demarteau_353,baur_demarteau,ewreport}
 The question addressed in the working group session was to consider 
the expected precision for $M_W$ at the VLHC in comparison to what will be 
available from competing facilities at VLHC turn-on. 
 For our estimates, we use 
 $\sqrt{s}= 100$ TeV, 
 $\Delta t = 16.7$ ns (the bunch spacing), 
 $\sigma_{tot} \simeq 120$ mb,
 and 20 interactions per crossing. 

For W events produced in a hadron collider environment there are
essentially only two observables that can be measured: {\it i)} the
lepton momentum, and {\it ii)} the transverse momentum of the recoil
system. The transverse momentum of the neutrino must be inferred from
these two observables.  The W boson mass can be extracted from either
the lepton transverse momentum distribution, or the transverse mass:
$M_T = \sqrt{2 p_T^e p_T^{\nu} (1-\cos \phi^{e\nu})}$, where
$\phi^{e\nu}$ is the angle between the electron and neutrino in the
transverse plane.

It is important to note that the following estimates necessitate a
large extrapolation from $\sqrt{s}= 1.8$ TeV to $\sqrt{s}= 100$
TeV.  For the W decays, the observed number distribution in
pseudorapidity ($\eta$) can be estimated by scaling results from the
CERN $Sp\bar{p}S$ and the Fermilab Tevatron.  The shoulder of the
pseudorapidity plateau is $\sim$ 3 for $\sqrt{s}= 630 \, \GeV$, and
$\sim$ 4 for $\sqrt{s}= 1.8$ TeV.  This yields an estimate in the
range of $\sim 5$ to 9 for a $\sqrt{s}= 100$ TeV VLHC.  Assuming
coverage out to $|\eta| \leq 4$, we obtain $\sim 1400$ charged tracks
in the detector calorimeter with which we must contend for the missing
$E_T$ calculation, ($\Emiss$).  Scaling the $\langle p_T \rangle$ up
to $\sqrt{s}= 100$ TeV we estimate $\langle p_T \rangle \simeq 865
\, MeV$ for minimum bias tracks.  Assuming $N_{ch}/N_{\gamma}=1$
yields an average $E_T$ flow of $2$ TeV in the detector.  Using
current $\Emiss_T$ resolutions of $\sim 4 - 5 \, \GeV$, we estimate
$\sigma(\Emiss_T) \simeq 25 - 30 \, \GeV$ for VLHC.

Two fundamental problems we encounter at a VLHC are multiple
interactions and pile-up.  Multiple interactions are produced in the
same crossing as the event triggered on.  The effects are
``instantaneous;" {\it i.e.}, the electronic signals are added to the
trigger signals and subjected to the same electronics.  Pile-up effects
are out-of-time signals from interactions in past and future buckets
caused by ``memory" of the electronics. Both cause a bias and affect
the resolution, but in different ways. The effect of pile-up is
strongly dependent on the electronics used in relation to the bunch
spacing.

The bottom line is the estimation of the total uncertainty on the W
mass, $\delta M_W$.  For a luminosity of 2~fb$^{-1}$, $\delta M_W$ is about 
20~MeV for both the transverse mass and lepton transverse momentum fits.
For an increased luminosity of $10fb^{-1}$, the transverse mass fit might
improve to $\delta M_W \sim 15 \, MeV$, with minimal improvement for
the determination from the lepton transverse momentum distribution.
It should be noted that these estimates have quite a few
caveats---additional study would be required before taking these
numbers as guaranteed predictions.  
In Table~\ref{tab:deltamw}, we compare these estimations
with the anticipated uncertainty from upcoming experiments.
Clearly the VLHC will not greatly improve the determination of
$M_W$.   The situation becomes more difficult when one insists that
the VLHC detectors be capable of {\em precisely} measuring  the relatively low 
energy leptons from the $M_W$ decay.

\begin{table}[htbp]
\begin{center}
\caption{
Anticipated limits on  $\delta M_W$ from present and future facilities.
(This compilation is 
taken from Ref.~\protect\cite{baur_demarteau}.)
}
\label{tab:deltamw}
\begin{tabular}{||l|c|r||} \hline \hline 
{\sc Facility }   &  $\delta M_W$ {\small ($MeV/c^2$)}  & ${\cal L}$  
\quad\quad  
\\ \hline
\hline NuTeV    & $\sim$ 100  &  --- \quad\quad  \\ \hline
HERA     & $\sim$ 60  &  150 $pb^{-1}$   \\ \hline
LEP2	    &  $\sim$ 35-45  &  $500 pb^{-1}$   \\ \hline
Tevatron & $\sim$ 55 & 1 $fb^{-1}$   \\ \hline
Tevatron & $\sim$ 18 & 10 $fb^{-1}$   \\ \hline
LHC      & $\lsim$ 15 & 10 $fb^{-1}$   \\ \hline
VLHC     & $\sim$ 20  & 1 $fb^{-1}$   \\ \hline
VLHC     & $\sim$ 15  & 10 $fb^{-1}$   \\ \hline  \hline
\end{tabular}
\end{center}
\end{table}

\section{The Top Quark\protect\footnote{%
\lowercase{\uppercase{B}ased on the presentation by 
\uppercase{E}rich \uppercase{V}arnes.}}
}

The mass of the recently discovered top quark is precisely determined
by the CDF and
\Dzero\  collaborations from $t \bar{t}$ production at the Tevatron.
 For the details of this discovery and measurement, we refer the
reader to Refs.~\cite{pbarp,observetop,tmass,dilepton}.

In Table~\ref{tab:top}, we display  the anticipated accuracy on the top 
quark mass at the Tevatron as estimated in the TeV2000 
report.\cite{tev2000}
Since this report, statistical techniques have been improved such that
one would expect a precision of $\delta m_t \sim  1.5$ GeV  with $10~
fb^{-1}$,  assuming other sources of systematics are negligible. 

Moving on to the LHC, the top production cross section is $\sim 100$ times 
greater than at TeV2000, so with a luminosity of $\sim 100 fb^{-1}/year$, 
we  expect $\sim 1000$ more top events after one LHC year. 
Assuming naively that the errors scale as $1/\sqrt{N}$
(where N is the number of events), we would obtain 
$\delta m_t \sim 50$ MeV. 

The challenges of the VLHC are quite similar to the LHC regarding this
measurement.  A precision measurement of the top quark mass at this
level (or better) places stringent demands on the jet calibration.
Even with large control samples of $Z$ + jets and $\gamma$ + jets,
uncertainties due to the ambiguous nature of jet definitions will
persist.  The large number of multiple interactions at LHC and VLHC
complicates this analysis (in a manner similar to that discussed for
the W boson mass measurement).  Therefore, in order to improve upon
existing measurements, the VLHC detectors will need to be extremely
well designed and understood---certainly a heroic task.

\begin{table}[htbp]
\begin{center}
\caption{
Anticipated accuracy on the top quark mass, as estimated 
by the TeV2000 report.\protect\cite{tev2000}
}
\label{tab:top}
\begin{tabular}{||c|r|r|r||} \hline \hline 
Source  & 70 $pb^{-1}$ & 1 $fb^{-1}$ & 10 $fb^{-1}$   \\ \hline \hline
Statistics & 25 & 6.2 & 2  \\ \hline
Jet Scale & 11 & 2.7 & 0.9   \\ \hline
Backgrounds & 4 & 1 & 0.3   \\ \hline  \hline
Total & 27.6 & 6.9 & 2.2   \\ \hline  \hline
\end{tabular}
\end{center}
\end{table}

\section{Probing a nonstandard Higgs boson at a VLHC\protect\footnote{%
\lowercase{\uppercase{B}ased on the presentation by 
\uppercase{V}assilis \uppercase{K}oulovassilopoulos.}}
}

 We have studied the potential of a VLHC to observe a nonstandard Higgs
 boson (i.e. a spin-0 isospin-0 particle with nonstandard couplings to
 weak gauge bosons and possibly fermions) and distinguish it from the
 Standard Model  Higgs boson.
 Results are presented for different options for the energy
 ($\sqrt{s}=50, 100, 200$~TeV) and luminosity (${\mathcal L}=10^{33}-10^{35}
 cm^{-2} s^{-1}$) and compared to those obtained for the LHC in  
Ref.~\cite{vki}.

 Our analysis is based on
 the gold-plated channel $H\rightarrow
 ZZ \rightarrow l^+l^-l^+l^-$
 and assumes cuts on the final-state leptons, which are given by
 $|\eta^l|<3, \,p_T^l > 0.5 \times 10^{-3} \sqrt{s}$.
 We studied Higgs masses in the range from 400 to 800 GeV (600-800~GeV
 for $\sqrt{s}=200$~TeV), where the lower limit is due to the cuts and
 the upper limit is theoretically motivated. 

The two relevant parameters
 that encode the deviations from the Standard Model (SM) are $\xi$ and $y_t$,
 the $HW^+W^- (HZZ$) and $Ht\bar{t}$ couplings relative to
 the SM respectively.
 We found that a nonstandard Higgs should be detected for practically all
 values of $\xi, y_t$ and ${\mathcal L}$ in the entire mass range studied,
 a situation which is not so clear for the LHC, particularly for
 the larger masses. 

A nonstandard Higgs boson can be distinguished
 from the SM one by a comparison of its width $\Gamma_H$ and the total
 cross-section. Due to theoretical uncertainties in the latter, we chose
 to use as a criterion only the measurement of the width. Following the
 procedure of ref.~\cite{vki} we quantified the statistical significance
 of a deviation from the SM prediction by constructing the probability
 density function according to which the possible measurements of the
 {\it Standard Model} width are distributed.
 Postulating that a nonstandard Higgs boson is ``distinguishable'' if its 
 width differs from the SM value by at
 least $3\sigma$, we were able to determine the precision with which
 the parameter $\xi$ can be measured at the LHC and a VLHC. This is
 summarized in Table~\ref{tab:sensitivity} for the case of $y_t=1$.
 We deduce that, for the purpose of precision
 measurements of the Higgs couplings, a lower energy VLHC with higher 
 luminosity is preferred to that of a higher energy with lower luminosity ---
 a conclusion that is due to the low-mass 
 character of the physics of interest.

Consequently, we find that for   Higgs masses in the
 range from 400 to 800 GeV,  the Higgs-Z-Z coupling
 can be measured to within a few percent at the VLHC, depending on the 
 precise mass and collider parameters.

 %
 \begin{table}
 \label{tab:sensitivity}
 \caption{
 Approximate sensitivity to the parameter $\xi$ at the LHC and the VLHC
 for various values of the luminosity and CM energy. The starred entries
 indicate that the value given applies only to $\xi > 1$, whereas for $\xi < 1$
 the sensitivity is substantially worse.}
 \begin{center}
 \begin{tabular}{|r|c|c|c|}     \hline \hline
 $\sqrt{s}$, ${\mathcal L}$ {\small (cm$^{-2}$ s$^{-1}$)} &
             \multicolumn{3}{c|}{Sensitivity to $\xi$} \\ \cline{2-4}
             & $m_H=400$~GeV & $600$ ~GeV & $800$~GeV \\ \hline
 14~TeV, $\;\; 10^{33}$ & 60\% $^*$  & ---   & ---   \\ \hline

 14~TeV, $\;\; 10^{34}$ & 20\% $^*$ & 40\% $^*$ & ---   \\ \hline

 50~TeV, $\;\; 10^{34}$ &  7\%  & 12\%  &  20\% \\ \hline

 50~TeV, $\;\; 10^{35}$ &  3\%  & 4\%   &  7\%  \\ \hline

 100~TeV, $\;\; 10^{34}$ & 6\%  & 8\%   &  12\%  \\ \hline

 100~TeV, $\;\; 10^{35}$ & 2-3\%  & 3\%   &  5\%   \\ \hline

 200~TeV, $\;\; 10^{33}$ &  --- & 25\%  &  30\%  \\ \hline

 200~TeV, $\;\; 10^{34}$ &  --- & 8\%   &  12\%   \\ \hline \hline
 \end{tabular}
 \end{center}
 \end{table}

\section{Supersymmetry\protect\footnote{%
\lowercase{\uppercase{B}ased on the presentation by 
\uppercase{J}oseph \uppercase{L}ykken.}}
}

Supersymmetry (SUSY) is a dominant framework for formulating physics
beyond the standard model in part due to the appealing
phenomenological and theoretical features.  SUSY is the only possible
extension of the spacetime symmetries of particle physics, SUSY easily
admits a massless spin-2 (graviton) field into the theory, and SUSY
appears to be a fundamental ingredient of superstring theory.  Given
the large number of excellent recent reviews and reports on 
SUSY,\cite{lykken,reportsup,lhcsusy} 
we will focus here on the issues directly related to the
VLHC.

One specific question which was addressed in the working group meeting was:
Is the VLHC a precision machine for standard weak-scale SUSY
with sparticle masses in the range 80 GeV to 1 TeV?
 Probably not, for the following reasons.
 \begin{itemize}
 \item
An order of magnitude increase in sparticle production rates 
will yield minimal gains, {\it except} for sparticles in the range
$\gsim$ 1 TeV. 
 \item 
Multiple interactions, degraded tracking, calibration, and b-tagging 
issues complicate reconstruction of the SUSY decay chains. 
 \end{itemize}
 \noindent
On the contrary, VLHC looks best if SUSY has some heavy surprises
in store such as $\gsim$ 1 TeV squarks, or $\sim$ 10 TeV SUSY messengers. 

 One example of a plausible SUSY scenario would be  heavy first and second 
generation squarks and sleptons (to suppress FCNC's) with a characteristic
mass in the range of $\sim$ 3 TeV. ({\it Cf.}, Ref.~\cite{lhcsusy}) 
While the gauginos and the third generation squarks and sleptons would be 
within reach of the LHC, investigation of 
 $\{ 
  \widetilde{u},
  \widetilde{d},
  \widetilde{e},
  \widetilde{\nu_e},
 \}$
and 
 $\{ 
  \widetilde{c},
  \widetilde{s},
  \widetilde{\mu},
  \widetilde{\nu_\mu}
 \}$
in the multi-TeV energy range would require a higher energy facility
such as the VLHC.  

 An estimate of the heavy squark signal over the weak-scale SUSY
background and conventional channels (such as $t\bar{t}$) indicates
that a VLHC can observe heavy quarks in the $\sim$ 3 TeV mass range;
such a heavy squark is difficult to reach at the LHC.  One might
expect on order of $10^3 - 10^4$ signal events/year. Of course,
background rejection is a serious outstanding question, and the
efficiently of b-tagging and high $p_t$ lepton rejection, for example,
are crucial to suppressing the backgrounds.

\section{Conclusions}

While these individual topics are diverse, there are some common themes
we can identify with respect to a VLHC machine. 
  First,  a very high energy hadron collider does not appear to be the
machine of choice for precision measurements in the energy range
$\lsim$500~GeV. The competition from Tevatron, HERA, LEP, and LHC are
formidable in this region. To obtain comparable precision, the
VLHC is handicapped by numerous factors including 
multiple interactions, 
large multiplicity, and large $\Emiss$. 
 Designing a  detector to operate in the VLHC environment
while achieving the precision of the lower energy competition
is a challenging task. 

In contrast, the strong suit of the VLHC is clearly its kinematic 
reach. Should there be unexpected sparticles in the $\gsim$~TeV 
range, the VLHC would prove useful in exploring this range. 
 Of course our intuition as to what might exist in the $\sim$10~TeV
regime is not as refined as the $\lsim$1~TeV regime which will
be explored in the near-future; however what
we discover in this energy range can provide important clues as
to where we should search with a VLHC.


\end{document}